\documentclass[11pt, letterpaper]{article}

\usepackage{ifthen}
\newcommand{\Bset}[1]{\setboolean{#1}{true}}
\newcommand{\Bunset}[1]{\setboolean{#1}{false}}

\usepackage{amssymb}
\usepackage{amsmath}  
\usepackage{amsthm}  
\usepackage{latexsym}
\usepackage{amsfonts}
\usepackage{psfrag}   
\usepackage{url}   

\usepackage[dvips]{graphicx}
\usepackage[usenames,dvipsnames]{color}

\newcommand{\Bif}[3]{\ifthenelse{\boolean{#1}}{#2}{#3}}

{\theoremstyle{remark}}

{\theoremstyle{definition}}

\newtheoremstyle{anystatement}{\topsep}{\topsep}{\itshape}{}{\bfseries}{.}{ }{\anystatementname}
{\theoremstyle{anystatement}}
\newcommand{\anystatementname}{}

\newcounter{tmp_id_cnt}
\newcommand{\nospell}[1]{#1}  

\newcommand{\mydef}[2]{\def#1{#2}}

\newcommand{\newident}[3][*]{\ifthenelse{\equal{*}{#1}}
{\newcommand{#2}[1][*]
{\ifthenelse{\equal{*}{##1}}
{\nospell{\mbox{\Ensuremath{{\mathit{#3}}}}}}
{\ifthenelse{\equal{b}{##1}}
{\nospell{\mbox{\Ensuremath{{\mathbf{#3}}}}}}
{#3}}}}
{\mydef{#2}{#3}}}

\newcommand{\newmat}[3][*]{\ifthenelse{\equal{*}{#1}}
{\newcommand{#2}[1][*]
{\ifthenelse{\equal{b}{##1}}
{\nospell{\mbox{\Ensuremath{\mathbf{#3}}}}}
{\ifthenelse{    \( \equal{*}{##1} \and \not \boolean{in_math_mode} \)
\or \( \not \equal{*}{##1} \and \boolean{in_math_mode} \)}
{\nospell{\mbox{\Ensuremath{#3}}}}
{#3}}}}
{\mydef{#2}{#3}}}

\newcommand{\newmatarg}[2]{ \newcommand{#1}[1][]
{\ifthenelse{\boolean{in_math_mode}}
{#2}
{\nospell{\mbox{\Ensuremath{#2}}}}}}

\newcommand{\newmatop}[2]{\mydef{#1}{\operatorname{#2}}}

\newcommand{\MyMakeTheoMacros}[3]{ \newcommand{#2}[2][]{\ifthenelse{\equal{}{##1}}
{\begin{#1} ##2 \end{#1}}
{\begin{#1}\label{##1} ##2\end{#1}}}
\newcommand{#3}[3][]{\ifthenelse{\equal{}{##1}}
{\begin{#1}{\e{##2}} ##3 \end{#1}}
{\begin{#1}{\e{##2}}\label{##1} ##3\end{#1}}}}

\newcommand{\MyMakeDupTheoMacros}[8]{ \MyMakeTheoMacros{#1}{#2}{#3}
\newcommand{#4}[3]{  \newcommand{##2}{##3}
\begin{#1}\label{##1} ##2\end{#1}}
\newcommand{#5}[4]{  \newcommand{##2}{##4}
\begin{#1}{\e{##3}}\label{##1} ##2\end{#1}}
\newcommand{#8}[2]{\def\my_tmp_id{my_tmp_id_\arabic{tmp_id_cnt}}
\newtheorem*{\my_tmp_id}{#7~\ref{##1}}
\begin{\my_tmp_id} ##2 \end{\my_tmp_id}\stepcounter{tmp_id_cnt}}
\newcommand{#6}[6]{   #2[##1]{##2}

##3
\prf[#7~\ref{##1}]{##6} \newcommand{##5}{}}}

\newcommand{\MyMakeRefMacros}[3]{\newcommand{#1}[2][]
{\ifthenelse{\equal{}{##1}}{#2~\ref{##2}}{#3~\ref{##1} and~\ref{##2}}}}

\newcommand{\MyMakeEqRefMacros}[3]{\newcommand{#1}[2][]
{\ifthenelse{\equal{}{##1}}{#2~\eqref{##2}}{#3~\eqref{##1} and~\eqref{##2}}}}

\newcommand{\abstr}[1]{	\begin{abstract}
#1
\end{abstract}}

\newcommand{\bibentry}[8]{\bibitem[\nospell{#8}]{#1} {\textup #3}. 
\ifthenelse{\equal{}{#6}}
{\newblock \textrm{#4.} \newblock {\em #5}, #7.}
{\newblock \textrm{#4.} \newblock {\em #5, #6}, #7.}}

\newcommand{\inputbib}{\bibentry{BCW98}{Buhrman, Cleve and Wigderson}{H. Buhrman, R. Cleve and A. Wigderson}{Quantum vs. Classical Communication and Computation}{Proceedings of the 30th Symposium on Theory of Computing}{pp. 63-68}{1998}{BCW98}

\bibentry{BCWW01}{Buhrman, Cleve, Watrous and de Wolf}{H. Buhrman, R. Cleve, J. Watrous and R. de Wolf}{Quantum Fingerprinting}{Physical Review Letters 87(16)}{article 167902}{2001}{BCWW01}

\bibentry{BJK04}{Bar-Yossef, Jayram and Kerenidis}{Z. Bar-Yossef, T. S. Jayram and I. Kerenidis}{Exponential Separation of Quantum and Classical One-Way Communication Complexity}{Proceedings of 36th Symposium on Theory of Computing}{pp. 128-137}{2004}{BJK04}

\bibentry{BS74}{Bondy and Simonovits}{J. Bondy and M. Simonovits}{Cycles of Even Length in Graphs}{Journal of Combinatorial Theory, Series B, 16}{pp. 87-105}{1974}{BS74}

\bibentry{CFL83}{Chandra, Furst and Lipton}{A. Chandra, M. Furst and R. Lipton}{Multi-party protocols}{Proceedings of the 15th Symposium on Theory of Computing}{pp. 94-99}{1983}{CFL83}

\bibentry{G07_C_I_C_R_a}{Gavinsky}{D. Gavinsky}{Classical Interaction Cannot Replace a Quantum Message}{Submitted}{http://arxiv.org/abs/quant-ph/0703215, http://eccc.hpi-web.de/eccc-reports/2007/TR07-058/}{2007}{G07}

\bibentry{GKKRW07}{Gavinsky, Kempe, Kerenidis, Raz and de Wolf}{D. Gavinsky, J. Kempe, I. Kerenidis, R. Raz and R. de Wolf}{Exponential Separations for One-Way Quantum Communication Complexity, with Applications to Cryptography}{Proceedings of the 39th Symposium on Theory of Computing}{}{2007}{GKKRW07}

\bibentry{K07}{Kerenidis}{I. Kerenidis}{Quantum Multiparty communication complexity and circuit lower bounds}{4th Annual Conference on Theory and Applications of Models of Computation}{}{2007}{K07}

\bibentry{LPS88}{Lubotzki, Phillips and Sarnak}{A. Lubotzki, R. Phillips and P. Sarnak}{Ramanujan Graphs}{Combinatorica 8}{pp. 261-277}{1988}{LPS88}

\bibentry{LUW95}{Lazebnik, Ustimenko and Woldar}{F. Lazebnik, V. Ustimenko and A. Woldar}{A New Series of Dense Graphs of High Girth}{Bulletin of AMS 32}{pp. 73-79}{1995}{LUW95}

\bibentry{N91}{Newman}{I. Newman}{Private vs. Common Random Bits in Communication Complexity}{Information Processing Letters 39(2)}{pp. 67-71}{1991}{N91}

\bibentry{R99}{Raz}{R. Raz}{Exponential Separation of Quantum and Classical Communication Complexity}{Proceedings of the 31st Symposium on Theory of Computing}{pp. 358-367}{1999}{R99}

\bibentry{Y79}{Yao}{A. C-C. Yao}{Some Complexity Questions Related to Distributed Computing}{Proceedings of the 11th Symposium on Theory of Computing}{pp. 209-213}{1979}{Y79}}

\newcommand{\bib}[1][]{   \begin{thebibliography}{ABCDEFG00}
\frenchspacing
\inputbib
#1
\end{thebibliography}}

\MyMakeTheoMacros{fact}{\fct}{\nfct}

\MyMakeRefMacros{\fctref}{Fact}{Facts}

\MyMakeDupTheoMacros{lemma}
{\lem}{\nlem}{\lemdup}{\nlemdup}{\lemapp}{Lemma}{\lemrep}

\MyMakeRefMacros{\lemref}{Lemma}{Lemmas}

\MyMakeDupTheoMacros{corollary}
{\crl}{\ncrl}{\crldup}{\ncrldup}{\crlapp}{Corollary}{\crlrep}

\MyMakeRefMacros{\crlref}{Corollary}{Corollaries}

\MyMakeTheoMacros{proposition}{\pr}{\npr}

\newtheorem*{prp*}{\e{Proposition}}

\MyMakeRefMacros{\prpref}{Proposition}{Propositions}

\MyMakeDupTheoMacros{claim}
{\clm}{\nclm}{\clmdup}{\nclmdup}{\clmapp}{Claim}{\clmrep}

\MyMakeRefMacros{\clmref}{Claim}{Claims}

\MyMakeDupTheoMacros{theorem}
{\theo}{\ntheo}{\theodup}{\ntheodup}{\theoapp}{Theorem}{\theorep}

\MyMakeRefMacros{\theoref}{Theorem}{Theorems}

\MyMakeTheoMacros{definition}{\defi}{\ndefi}

\MyMakeRefMacros{\defiref}{Definition}{Definitions}

\MyMakeTheoMacros{problem}{\prob}{\nprob}

\MyMakeRefMacros{\probref}{Problem}{Problems}

\MyMakeTheoMacros{protocol}{\prot}{\nprot}

\MyMakeRefMacros{\protref}{Protocol}{Protocols}

\providecommand{\qedsymbol}{\square}

\newcommand{\prf}[2][]{\ifthenelse{\equal{}{#1}}
{\begin{proof}\renewcommand{\qedsymbol}{$\blacksquare$} #2 \end{proof}}
{\begin{proof}[Proof of #1]
\renewcommand{\qedsymbol}{$\blacksquare_{\mbox{\it{\scriptsize{#1}}}}$}
#2 \end{proof}}}

\newcommand{\sect}[2][]{\ifthenelse{\equal{}{#1}}
{\section{#2}}
{\section{#2}\label{#1}}}

\newcommand{\ssect}[2][]{\ifthenelse{\equal{}{#1}}
{\subsection{#2}}
{\subsection{#2}\label{#1}}}

\MyMakeRefMacros{\chref}{Chapter}{Chapters}

\MyMakeRefMacros{\sref}{Section}{Sections}

\MyMakeRefMacros{\ssref}{Subsection}{Subsections}

\MyMakeRefMacros{\sssref}{Subsection}{Subsections}

\newboolean{in_math_mode}
\setboolean{in_math_mode}{false}

\newcommand{\IfMathMode}{\Bif{in_math_mode}}

\newcommand{\MathModeOn}{\Bset{in_math_mode}}

\newcommand{\MathModeOff}{\Bunset{in_math_mode}}

\newcommand{\Ensuremath}[1]{\IfMathMode  
{\ensuremath{#1}}
{\MathModeOn\ensuremath{#1}\MathModeOff}}

\newcommand{\fbr}[1]{\IfMathMode  
{$#1$}                             
{\MathModeOn$#1$\MathModeOff}}

\newcommand{\fnbr}[1]{\mbox{\fbr{#1}}}  

\newcommand{\fla}[2][*]{\ifthenelse{\equal{}{#1}}{\fbr{#2}}{\fnbr{#2}}}

\newcommand{\mat}[2][]{\ifthenelse{\equal{}{#1}}  
{\begin{displaymath} \MathModeOn
#2
\MathModeOff \end{displaymath}    }{  \begin{equation} \MathModeOn \label{#1}
#2
\MathModeOff \end{equation}}}

\newcommand{\f}{\fla}

\newcommand{\m}{\mat}

\MyMakeEqRefMacros{\equref}{Equation}{Equations}

\MyMakeEqRefMacros{\expref}{Expression}{Expressions}

\MyMakeEqRefMacros{\inequref}{Inequality}{Inequalities}

\MyMakeRefMacros{\figref}{Figure}{Figures}

\providecommand{\middle}{\big}

\newmatop{\mymod}{mod}  
\newmatop{\poly}{poly}
\newmatop{\sign}{sign}
\newmatop{\tr}{tr}
\newmatop{\supp}{supp}   
\newmatop{\argmax}{argmax}
\newmatop{\argmin}{argmin}
\newmatop{\posmin}{posmin}
\newmatop{\negmin}{negmin}
\newmatop{\posmax}{posmax}
\newmatop{\negmax}{negmax}
\newmatop{\argmaxmin}{argmax(min)}

\newcommand{\ord}[1][]{\nospell{\ifthenelse{\equal{}{#1}}
{\mbox{'th}}
{\ifthenelse{\equal{1}{#1}}{$1$\mbox{'st}}{\ifthenelse{\equal{2}{#1}}{$2$\mbox{'nd}}{\ifthenelse{\equal{3}{#1}}{$3$\mbox{'rd}}{\fla{#1}\mbox{'th}}}}}}}

\newmat[]{\llp}{\left(}
\newmat[]{\rrp}{\right)}

\newmat[]{\llc}{\left\lceil}
\newmat[]{\rrc}{\right\rceil}

\newmat[]{\tm}{\cdot}
\newmat[]{\xor}{\oplus}
\newmat[]{\sbseq}{\subseteq}
\newmat[]{\eps}{\varepsilon}
\newmat[]{\deq}{\stackrel{\textrm{def}}{=}}

\newmat[]{\ds}{\dots}
\newmat[]{\dc}{,\ds,}
\newmat[]{\dtm}{\times\ds\times}
\newmat[]{\dcirc}{\circ\ds\circ}

\newcommand{\enum}[1]{\begin{enumerate} #1 \end{enumerate}}

\newcommand{\itemi}[2][]{\ifthenelse{\equal{}{#1}}
{\begin{itemize} #2 \end{itemize}}
{\begin{itemize}[#1] #2 \end{itemize}}}

\newcommand{\ea}{\il{et al.}\ }		

\newcommand{\fr}[3][*]{ \ifthenelse{\equal{*}{#1}}              
{\frac{#2}{#3}}{}
\ifthenelse{\equal{}{#1}}               
{\left.#2\middle/#3\right.}{}
\ifthenelse{\equal{b_}{#1}}             
{\left.\left(#2\right)\middle/#3\right.}{}
\ifthenelse{\equal{_b}{#1}}             
{\left.#2\middle/\left(#3\right)\right.}{}
\ifthenelse{\equal{bb}{#1}}             
{\left.\left(#2\right)\middle/\left(#3\right)\right.}{}}

\newcommand{\set}[2][]{\ifthenelse{\equal{}{#1}}
{\Ensuremath{\left\{#2\right\}}}
{\Ensuremath{\left\{#2\middle|\vphantom{|_1^1}#1\right\}}}}

\newcommand{\newfunction}[2]{ \newcommand{#1}[2][*]{\ifthenelse{\equal{*}{##1}}
{\Ensuremath{#2\llp ##2 \rrp}}
{#2(##2)}}}

\newfunction{\asO}{O}
\newfunction{\astO}{\tilde O}
\newfunction{\aso}{o}
\newfunction{\asOm}{\Omega}
\newfunction{\astOm}{\tilde \Omega}
\newfunction{\asom}{\omega}
\newfunction{\asT}{\Theta}

\newmat[]{\01}{\set{0,1}}

\newcommand{\sz}[2][]{\ifthenelse{\equal{}{#1}}
{\Ensuremath{\left|#2\right|}}
{\Ensuremath{\left|#2\right|_{#1}}}}

\newcommand{\ceil}[2][*]{\ifthenelse{\equal{}{#1}}
{\lceil #2 \rceil}
{\llc #2 \rrc}}

\newcommand{\fn}{\footnote}

\newcommand{\e}{\emph}

\newcommand{\bl}[1]{{\bf #1}} 

\newcommand{\il}[1]{{\it #1}}

\usepackage{hyphenat}  

\providecommand{\middle}{\big}

\providecommand{\ds}{\dots}
\providecommand{\dc}{,\ds,}
\providecommand{\dtm}{\times\ds\times}
\providecommand{\dcirc}{\circ\ds\circ}
\providecommand{\ea}{\il{et al.}\ }

\newcommand{\A}{{\bf A}}
\newcommand{\B}{{\bf B}}
\newcommand{\C}{{\bf C}}

\newmatarg{\HMP}{\ifthenelse{\equal{}{#1}}
{HMP^{(n,k)}}
{HMP_{#1}^{(n,k)}}}

\title{Exponential Separation of Quantum and Classical\\
Non-Interactive Multi-Party Communication Complexity}

\author{Dmitry Gavinsky
\thanks{David R. Cheriton School of Computer Science and Institute for Quantum Computing, University of Waterloo}
\and Pavel Pudl\'ak
\thanks{Institute of Mathematics, Academy of Sciences, \v Zitna 25,
Praha 1, Czech Republic. Supported by grant A1019401} }

\begin{document}

\maketitle

\thispagestyle{empty}

\abstr{We give the first exponential separation between quantum and classical multi-party communication complexity in the (non-interactive) one-way and simultaneous message passing settings.

For every $k$, we demonstrate a relational communication problem between $k$ parties that can be solved exactly by a quantum simultaneous message passing protocol of cost \asO{\log n} and requires protocols of cost $n^{c/k^2}$, where $c>0$ is a constant, in the classical non-interactive one-way message passing model with shared randomness and bounded error.
Thus our separation of corresponding communication classes is superpolynomial as long as $k=\aso{\sqrt{\fr{\log n}{\log\log n}}}$ and exponential for $k=\asO1$.}

\setcounter{page}{1}

\sect{Introduction}

In this paper we study quantum computation from the perspective of
communication complexity. In the two-party model, defined by
Yao~\cite{Y79}, two players are to compute a function of two variables
$x$ and $y$, each knowing only one of the variables. The complexity
measure is the number of bits they need to exchange in the worst case.
In general players may use shared randomness. 

An important generalization is the multi-party communication
complexity. In this paper we shall study the most important version of
multi-party communication, the {\em number on a forehead}
model, defined by Chandra, Furst and Lipton~\cite{CFL83}.  In
this case a function of $k$ variables $x_1,\dots,x_k$ is computed by
$k$ players each knowing $k-1$ of the variables, namely player $i$
knows $x_1,\dots,x_{i-1},x_{i+1},\dots,x_k$.  The definition naturally
generalizes to the case of \e{relational problems} (or \e{relations}),
where for a given input there may be several correct outputs, or none.

Obviously, the case of $k=2$ players coincides with the standard
two-party model.  On the other hand, proving lower bounds for $k>2$
players is usually much harder, since they share some common
information.

It has been established that quantum communication is exponentially
more efficient in a number of versions of the \e{two-party model}, see
\cite{BCW98,R99,BCWW01,BJK04,GKKRW07,G07_C_I_C_R_a}. 
The model of \e{multi-party quantum communication} has been defined by Kerenidis~\cite{K07}.
In this paper we shall give the first exponential separation between quantum and classical multi-party communication complexity. 

We shall consider two versions of the \e{non-interactive} model.
In the \e{one-way message passing} model, the first $k-1$ players send one message each to the \ord[k] player.
The latter is supposed to give an answer based on the received messages and his portion of input.\fn
{Note that we consider the \e{non-interactive} one-way model, as opposed to the following possible scenario:\ Alice, Bob and Charlie use a protocol, where Alice sends a message to Bob, \e{after that} Bob sends a message to Charlie, who in turn produces an answer.
Cf.\ \sref{sec_opr}.}
In the \e{simultaneous message passing} (\e{SMP}) model, each of the $k$ players sends a single
message to a \e{referee}, who is supposed to answer based solely on the received messages.
Of course, the model of SMP is, in general, weaker than the one-way model, because the answering side (the referee) does not have free access to any piece of input.

We shall show an exponential separation between quantum and
classical probabilistic communication complexity in these
models.
Specifically, for every $k$ we construct a relation and a quantum protocol that uses \asO{\log n} quantum bits to solve the
problem \e{exactly} in the SMP model, and prove that its classical probabilistic communication
complexity is $n^{\Omega(1)}$, even if we allow bounded error.

The exponent in the lower bound decreases with the number of players as $1/k^2$, as long as $k<c_1\tm\sqrt{\log n}$, $c_1>0$.
Thus we get superpolynomial separation as long as the number of players is in \aso{\fr{\log n}{\log\log n}}, and exponential separation for constant number of players.
The lower bound still holds for the stronger model of classical non-interactive one-way communication, even if we allow public randomness (our protocol, like any exact protocol, does not need randomness).

\sect{Definitions and notation} 
We write $\log$ to denote the logarithmic function with base $2$.

Let $P\sbseq X_1\dtm X_k\times Z$, where $X_1\dc X_k\sbseq\01^n$ are the domains of \e{arguments} and $Z\sbseq\01^*$ is the \e{range} of the relation $P$.
We say that a \f k-party protocol $S$ solves $P$ with error bounded by $\eps$ if the following holds\itemi{ \item $S$ describes behavior (i.e., which message is sent in every possible case, who produces the answer, and when) by the $k$ players and (optionally) the referee.
\item If $x_1\in X_1\dc x_k\in X_k$ are arguments to $P$ such that the set \set[(x_1\dc x_k,z)\in P]{z\in Z} is not empty, and for $1\le i\le k$, the \ord[i] player is given the values of $x_1\dc x_{i-1},$ $x_{i+1}\dc x_k$, then the answer $z_0\in Z$ produced by $S$ is correct (i.e., $(x_1\dc x_k,z_0)\in P$) with probability at least $1-\eps$.}
The \e{cost} of $S$ is the maximum possible total number of bits (classical or quantum) communicated before an answer is produced.

We call a protocol \e{(non-interactive) one-way} if the first $k-1$ players send at most one message each to the \ord[k] player, and no other communication occurs; after that the player $k$ produces an answer.

We call a protocol \e{simultaneous message passing} (\e{SMP}) if each player sends at most one message to the referee, and no other communication occurs; after that the referee produces an answer.

We shall consider the following problem. The input consists of $k-1$
indices $\alpha_1,\dots,\alpha_{k-1}$ and a string $c$ of $n$ bits. The
indices jointly determine a matching on $\{ 1,\dots,n\}$. The parts of
the input are distributed among the $k$ players as usual, thus each of
the first $k-1$ players knows $k-2$ indices and $c$, whereas player
$k$ knows all the indices, but does not know $c$. The goal is to
compute $i_1,i_2,c_{i_1}\xor c_{i_2}$, where $(i_1,i_2)$ is an edge of
the matching determined by the indices.

Now we shall describe the problem formally. Let $M_n=(m^{(n)}_i)_{i=1}^{t(n)}$ be
a family of $t(n)$ edge-disjoint perfect matchings over $n$ nodes and
$M=\{ M_n\}_{n\in N}$. Since the matchings $m_i^{(n)}$ are disjoint,
we have $t(n)\leq n$. 

\defi{Let $2<k<\log(t(n))$, such that $\log(t(n))$ is a multiple of $k-1$, and let $r(n)\deq\fr{\log(t(n))}{k-1}$.
Let $c\in\01^n$ and $\alpha_1\dc \alpha_{k-1}\in\01^{r(n)}$.  
Denote by $\circ$
concatenation of binary strings; interpret
$\alpha_1\dcirc\alpha_{k-1}$ as an integer between $1$ and
$2^{(k-1)r(n)}$.
Then $\big(\alpha_1\dc \alpha_{k-1},c,(i_1,i_2,c_{i_1}\xor c_{i_2})\big)\in\HMP[M]$ if $\alpha_1\dcirc\alpha_{k-1}\le t(n)$ and $(i_1,i_2)\in m^{(n)}_{\alpha_1\dcirc\alpha_{k-1}}$.}

\sect{A quantum protocol for \HMP[M]}

\pr[pro_prot]{  There exists a quantum \f k-party SMP communication protocol that
exactly solves \HMP[M] using \asO{\log n} quantum bits for any $M$.}

\prf{Recall that $\sz{M_n}\le n$.
Consider the following protocol.
\itemi{ \item Player $1$ sends the quantum state
\m{\frac 1{\sqrt{n}}\sum_{i=1}^n (-1)^{c_i}|i\rangle}
to the referee, which is $\ceil{\log n}$ quantum bits.
\item Player $k$ determines (based on his input) the matching $m^{(n)}_{\alpha_1\dcirc\alpha_{k-1}}$ and sends its index to the referee.
That costs $\ceil{\log\llp\sz{M_n}\rrp}\le\ceil{\log n}$ (classical) bits.
\item The referee performs a projective measurement in the orthogonal basis
\m{\set[(i_1,i_2)\in m^{(n)}_{\alpha_1\dcirc\alpha_{k-1}}]{|i_1\rangle \pm |i_2\rangle}.}
When he obtains $|i_1\rangle +|i_2\rangle$ ($|i_1\rangle -|i_2\rangle$), then $c_{i_1}\xor c_{i_2}=0$ ($c_{i_1}\xor c_{i_2}=1$, respectively), and the answer is produced accordingly.}}

\sect{Lower bound for solving \HMP[M] in the classical model}
In this section we show that for some choice of $M$ solving \HMP[M] in the classical non-interactive one-way model allowing bounded error and shared randomness is expensive.

\lem[lem_det]{For every $\eps_1,\eps_2>0$ there exists a constant $C$ such that for
every $k$ and $n$ the following holds.
If a one-way \f k-party protocol $S$ of cost $l$ uses shared randomness and solves \HMP[M] with error bounded by $\eps_1$, then for any constant $\eps_2$ there exists a protocol $S'$, satisfying\itemi{ \item $S'$ solves \HMP[M] with error at most $\eps_1+\eps_2$, for every possible input;
\item $S'$ has communication cost at most $(\log n)^{C}\tm l$;
\item the senders (first $k-1$ players) are deterministic, the recipient (\ord[k] player) may use private randomness.}}

Intuitively, the lemma says that we can ``partially derandomize'' the
protocol $S$, preserving its correctness in distribution-free setting
(which would not be the case if we simply applied the Min-Max Theorem).

\prf[\lemref{lem_det}]{We shall apply a result of Newman~\cite{N91},
Proposition~1.1, that shows that the number of shared random bits
can be reduced to constant time the logarithm of the input
size. More precisely, there exists absolute constants $c_1,c_2$ such
that for every $\eps,\delta>0$, every protocol that uses $l$
communication bits and solves the problem with error $\eps$ can
be replaced by a protocol that uses at most $c_1 l$ communication
bits, $c_2 \log n$ random bits (where $n$ is the input size) and has error
$\leq \eps+\delta$. This theorem was proved only for the two party
communication complexity, however the proof is completely general
and works for any number of parties and essentially any special way
of communication. In particular, the constants $c_1,c_2$ do not
depend on the number of players.

Let $S$ be as suggested by the lemma.  Applying Newman's result we
conclude that some protocol $S_1$ uses only \asO{\log n} shared
random bits (and no private randomness), has cost $l$ and solves
\HMP[M] with error at most $\eps_1+\eps_2/2$.

Then there exists a protocol $S_2$ of cost $l+\asO{\log n}$, solving
the problem with the same error, but with public randomness shared
only between the first $k-1$ players (in $S_2$, one of the senders
appends to his message the content of the random string).

Now let us consider the following communication task between the
first $k-1$ players: They receive same input as the first $k-1$
players of $S_2$, and their goal is to produce messages which would,
according to $S_2$, cause the \ord[k] player to produce a correct
answer (the \ord[k] player is deterministic in $S_2$, thus the
problem is well-defined).  Observe that all $k-1$ players share the
knowledge of $c$, so only the part of input which determines $m\in
M_n$ is not available to each player separately.  Note also that the
protocol $S_2$ (``restricted'' to the first $k-1$ players) solves
this problem with error at most $\eps_1+\eps_2/2$.  Recall that
$\sz{M_n}\le n$.  We apply Newman's theorem for the second time,
concluding that there exists a protocol allowing the first $k-1$
players to accomplish their task with success probability at least
$1-\eps_1-\eps_2$ using
$\asO{\log\llp\log(\sz{M_n})\rrp}=\asO{\log\log n}$ shared
random bits.

Therefore, there exists a protocol $S_3$ of cost $l+\asO{\log n}$ that solves \HMP[M] with error at most $\eps_1+\eps_2$, uses \asO{\log\log n} random bits shared between the first $k-1$ players and no other randomness.

Finally, let us derandomize the first $k-1$ players of $S_3$.
They share \asO{\log\log n} random bits, they can take one of $(\log n)^{\asO1}$ possible values.
Define $S'$ as follows.
Let each of the first $k-1$ players send the sequence of messages which he would send, according to $S_3$, with respect to all possible values of random bits.
The recipient (\ord[k] player) randomly chooses one possible value of the random bits, considers only those parts of the messages which correspond to that value and acts according to $S_3$.
This protocol satisfies the requirement of the lemma.}

\ssect{On the families of perfect matchings}
We will construct a family $M'$ of perfect matchings that makes \HMP[M'] hard for the classical model.
Our construction is based on some results in extremal graph theory
concerning the number of edges in graphs with forbidden subgraphs. Let 
\[
ex(n,\{ G_1,\dots,G_j\}),
\] 
denote the maximal number of edges that a graph on $n$ vertices can
have without containing any of the graphs $G_1,\dots,G_j$ as (not
necessarily induced) graphs. These numbers have been studies
especially for cycles $C_d$. By a result of Bondy and Simonovits~\cite{BS74}
\[
ex(n,\{ C_{2d}\})\leq 90 d n^{1+1/d}.
\]
Lower bounds of the form
\[
ex(n,\{C_3,C_4,\dots, C_{2d}\})=\Omega( n^{1+2/(3d+\nu)}),
\]
have been shown by Lubotzki, Phillips and Sarnak (with $\nu=3$)
\cite{LPS88} and Lazebnik, Ustimenko and Woldar \cite{LUW95} (with $\nu=-2$
and $\nu=-3$, depending on the parity of $\nu$). These bounds were 
obtained using explicit constructions. These constructions are,
moreover, bipartite regular graphs (i.e., the degrees of all vertices
are equal). 

Our main combinatorial lemma is an immediate corollary of these results.

\lem[l-ex]{For every $d$ and every prime power $t$ there exist a number $n$
and a bipartite $t$-regular graph $G_{n,d}$ on $n$ vertices such that
\begin{enumerate}
\item $n\leq t^{\frac 32 d}$,
\item $G_{n,d}$ can be decomposed into $t$ disjoint perfect matchings,
\item every set of edges $E$ spans at least $|E|^{1-1/(d+1)}/90d$ vertices.
\end{enumerate}} \prf{In \cite{LUW95} Lazebnik \ea constructed $t$-regular
bipartite graphs satisfying the first condition and such that they
do not contain $C_{2d}$. It is well-known that regular bipartite
graphs can be decomposed into edge-disjoint perfect matchings.
(Namely, one can easily check that the assumption of Hall's theorem
is satisfied, hence the graph contains a perfect matching. If we
delete the edges of this matching the remaining graph is still
bipartite and regular.) The condition about forbidden cycles implies
3., according to the result of Bondy and 
Simonovits~\cite{BS74}.}

For our lower bound on $k$-party communication complexity we need the
number of matchings be of the form $t=2^{r(k-1)}$, therefore we shall
consider only powers of 2. For a $t$ of this form, let $n$ and
$G_{n,2k}$ be the number and the graph from the lemma.  We shall
define a family of perfect disjoint matchings
$M_{n,k}=(m^{(n,k)}_i)_{i=1}^{t}$ to be the perfect matchings of
$G_{n,2k}$. 
It would be more natural to parametrize this family by $r$
and $k$, since for each pair $r$ and $k$ we have one such family of
matchings (with $t=2^{r(k-1)}$ and $n\leq 2^{3k(k-1)r}$). We use $n$
instead of $r$, since it indicates the size of inputs. For future
reference we note that $$t\geq n^{1/3k}.$$

\ssect{Lower bound for \HMP[M_{n,k}]} First we recall some properties
of the {\em mutual information} of random variables that we shall need
in the proof. Let $\bf X$ and $\bf Y$ be random variables, then we
define their mutual information by
\[
I({\bf X;Y}) = H({\bf X})+ H({\bf Y})- H({\bf X,Y})=  H({\bf X})- H({\bf X|Y}),
\]
where $H$ is entropy. We shall need the following facts:
\begin{enumerate}
\item $H({\bf X|Y})=\sum_y H({\bf X|Y}=y)\cdot Pr({\bf Y}=y).$
\item $I({\bf X;Y|Z})=\sum_z I({\bf X;Y|Z}=z)\cdot Pr({\bf Z}=z)$.
\item If ${\bf Y}_1,\dots,{\bf Y}_n$ are independent, then
\[
I({\bf X};{\bf Y}_1,\dots,{\bf Y}_n)\geq \sum_j I({\bf X};{\bf Y}_j).
\]
\end{enumerate}
The first fact follows from the definition by direct computation. The
second one is a consequence of the first one.  To prove the third
fact, write
\[
I({\bf X};{\bf Y}_1,\dots,{\bf Y}_n)=
H({\bf Y}_1,\dots,{\bf Y}_n)-H({\bf Y}_1,\dots,{\bf Y}_n|{\bf X}).
\]
Then express the first term as the sum of entropies and apply the
subadditivity of entropy to the second term.

We shall also use Markov's
inequality in the following form. If $0\leq {\bf X}\leq \beta$ and
$0\leq\alpha<\beta$, then
\[
Pr({\bf X}\geq \alpha)\geq \frac{E({\bf X})-\alpha}{\beta-\alpha},
\]
where $E$ denotes expectation. 

\theo[theo_bound]{For every $\eps>0$ there exists a constant $\gamma>0$ such that for every $k\ge2$, $k=\aso{\sqrt{\fr{\log n}{\log\log n}}}$, any non-interactive one-way protocol solving \HMP[M_{n,k}] with error $\fr12-\eps$ has to communicate at least 
\(
n^{\gamma/k^2}
\)
bits of information.}

\prf[\theoref{theo_bound}]{ Let $n$ be fixed.
Recall that $t\geq n^{1/{3k}}$ and  $r=\log t/(k-1)$.

Let $S$ be a \f k-party protocol of cost $l$, satisfying the theorem
requirement.  Let $1/2-\eps$ be the guaranteed upper bound on the
error probability of $S$.  Let $S'$ be another protocol with error
at most $1/2-\eps/2$, as guaranteed by \lemref{lem_det}.

Let $c\in\01^n$ and consider the $2^{(k-2)r}$ inputs of the form
\[
(\alpha_1,\dots,\alpha_{k-2},\sum_{i=1}^{k-2}\alpha_i,c).
\]
We interpret strings $\alpha_i$ in the sum as vectors in
$GF_2^{r}$.  Notice that for {\em any} subset of $k-2$ of the first
$k-1$ coordinates we get all $2^{(k-2)r}$ values of the
$k-2$-tuples of strings. 
Let $w$ be concatenation of the strings of
messages that the first $k-1$ players send to player $k$ for these
inputs, assuming they are using the protocol $S'$. Since each of these
players can see only $k-2$ coordinates from the first $k-1$
coordinates, the string $w$ encodes all messages that they ever send
for the given string $c$.

Recall that by \lemref{lem_det} the first $k-1$ players in $S'$ are
deterministic, and therefore for every tuple
$(\alpha_1\dc\alpha_{k-1})$ we can, using $w$, prepare the $k-1$
messages which are received by the \ord[k] player when the input is
$(\alpha_1\dc\alpha_{k-1},c)$ (and the input of the \ord[k] player
himself is $\alpha_1\dc\alpha_{k-1}$, i.e., the encoding of the
matching).  Consequently, for each matching $m\in M_{n,k}$ it is
possible to obtain, using the information contained in $w$, a triple
$(i_1,i_2,e)$, such that $(i_1,i_2)\in m$ and $e=c_{i_1}\oplus
c_{i_2}$ with probability at least $\fr12+\fr{\eps}2$.  Consider the
following algorithm that constructs a string of pairs of indices $A$
and a string of bits $B$ using $w$ as the input.
\enum{ \item Let $A$ and $B$ be empty strings initially.
\item Let $m\in M_{n,k}$, such that every edge in $m$ has at
most one endpoint in the support of the pairs of $A$.  If no such
matching exists, \bl{halt}.
\item Using $w$, get a triple $(i_1,i_2,e)$, such that $(i_1,i_2)\in
m$ and $e=c_{i_1}\oplus c_{i_2}$ with probability at least
$\fr12+\fr{\eps}2$ (this is the answer that player $k$ produces
given the messages of the players $1,\dots,k-1$ and the matching $m$).
\item Let $A:=A\circ \set{i_1,i_2}$ and $B:=B\circ e$.
\item Return to Step $2$.}

It follows from the properties of $M_{n,k}$ that if we take one
edge from each $m'\in M_{n,k}$ then at least $t^{1-\frac{1}{2k+1}}/180k$
vertices are touched by those edges.  Therefore, as long as
$\sz{A}<t^{1-\frac{1}{2k+1}}/180k$, it is possible to find $m\in M_{n,k}$
which satisfies the requirement of Step $2$.  On the other hand, each
iteration of the algorithm adds at most $2$ new elements to $A$,
therefore the algorithm always performs at least $\fr12 t^{1-\frac{1}{2k+1}}/180k$
iterations. Let $s$ be the least number of the steps of the
algorithm (the lengths of the strings $A$ and $B$), thus
\begin{equation}\label{e-a}
s\geq  t^{1-\frac{1}{2k+1}}/360k.
\end{equation}

We shall show that the mutual information between $(A,B)$ and $c$ is
$\Omega(s)$. Intuitively it seems clear, because the pairs $A$ chosen
by the algorithm form a tree, thus the bits $c_{i_1}\oplus c_{i_2}$
are independent, and $e=c_{i_1}\oplus c_{i_2}$ with probability at
least $\fr12+\fr{\eps}2$. Notice, however, that the pairs in $A$ and
the bits in $B$ are not independent, as we are choosing next matching
according to the outcome of the previous stage. Therefore a formal
proof of this fact is needed.

Consider the following random variables.
\begin{itemize}
\item ${\bf C}$ -- the uniform distribution on the strings $c\in\{  0,1\}^n$.
\item $\bf W$ -- the distribution on the strings $w$ when the uniform
distribution on the strings $c$ is uniform; thus $\bf W$ is a
function of ${\bf C}$.
\item $\bf A$ and $\bf B$ -- the distribution on the strings produced
by the above algorithm when the distribution on strings $c$ is
uniform; these random variables can be viewed as functions of $\bf
W$ and some random variable independent of ${\bf C}$ (the random
bits of player $k$).
\end{itemize}
The assumption about $\A$ and $\B$ can be stated as follows. For
every $j=1\dc s$,
\[
Pr(\B_j=\C_{(\A_j)_1}\oplus\C_{(\A_j)_2})\geq \frac 12 + \frac{\eps}2.
\]

Our proof of the theorem is based on estimating $I({\bf A,B;C})$
in two ways. The upper bound is easy: 
\begin{equation}\label{e0}
I({\bf A,B;C})\leq I({\bf W;C}) \leq H({\bf W})\leq |{\bf W}|,
\end{equation}
since $\bf A,B$ are functions of the random variable $\bf W$ and a
random variable independent of $\bf C$.

To prove a lower bound on $I({\bf A,B;C})$, we consider two cases. 

\bigskip
(1) $H(\C|\A)< n-\xi s$. Here $\xi>0$ is a sufficiently small
fraction of $\eps$ that will be specified later. In this case
\[
I(\A,\B;\C)\geq I(\A;\C)=H(\C)-H(\C|\A)> n -(n-\xi s) = \xi s.
\]

\bigskip
(2)  $H(\C|\A)\geq n-\xi s$. Observe that according to
the chaining rule
\[
I({\bf A,B;C})=I({\bf A;C})+I({\bf B;C}|{\bf A})\geq I({\bf B;C}|{\bf A}).
\]
Hence it suffices to estimate $I({\bf B;C}|{\bf A})$. 

Let $\bf D$ be the
random variable whose value is the number of indices $j$ such that 
$\B_j=\C_{(\A_j)_1}\oplus\C_{(\A_j)_2}$.
The assumption about the
correctness of the protocol implies that
\[
E({\bf D})\geq (1/2 +\eps/2)s,
\]
Consider the mapping
\[
A\mapsto E({\bf D}|{\bf A}=A)
\]
as a random variable.
Let 
\begin{equation}\label{e2}
\Delta_1(A)\equiv_{df}E({\bf D}|{\bf A}=A)\geq (1/2+\eps/4)s.
\end{equation}
By Markov's inequality,
\begin{equation}\label{e1}
Pr(\Delta_1(\A))=
\sum_{A;\Delta_1(A)} 
Pr({\bf A}=A)\geq \frac{\eps/2}{1-\eps/2}\geq \eps/2.
\end{equation}

Let 
\[
\Delta_2(A)\equiv_{df} H(\C|\A=A)\geq n-\frac 4{\eps}\xi s.
\]
In a similar fashion, we get
\[
Pr(\Delta_2(\A))\geq 1 - \frac{\eps}4.
\]
Whence
\[
Pr(\Delta_2(A)\wedge\Delta_2(A)\geq \eps/2 -\eps/4=\eps/4.
\]

By (\ref{e1}),
\begin{equation}\label{e-b}
I({\bf B;C|A})=\sum_a I({\bf B;C|A}=A)\cdot Pr({\bf A}=A)\geq
\frac{\eps}4\cdot \min_{A;\Delta_1(A)\wedge\Delta_2(A)} I({\bf B;C|A}=A).
\end{equation}
So it remains to estimate $I({\bf B;C|A}=A)$ for $A$ satisfying
$\Delta_1\wedge\Delta_2$. 

Let such an $A$ be fixed. Let ${\bf C'}_j$ be the random
variables defined as follows. For $j=1,\dots,s$, let 
${\bf C'}_j={\bf  C}_{i_1}\oplus  {\bf C}_{i_2}$ where $(i_1,i_2)$ is
the $j$-th pair of $a$. For the rest of indices $j$, we set each ${\bf
C'}_j$ to be equal to ${\bf C}_i$ for some $i$, so that the
variables ${\bf C'}_j$ are independent. Then the information contained
in $\bf C$ and $\bf C'$ is the same, so we can replace the first by
the second. Thus
\[
I({\bf B;C|A}=A)=I({\bf B;C'|A}=A)\geq 
\sum_j I({\bf B};{\bf C'}_j|{\bf A}=A)\geq 
\]\[
\sum_{j=1}^s I({\bf B};{\bf C'}_j|{\bf A}=A)\geq
\sum_{j=1}^s I({\bf B}_j;{\bf C'}_j|{\bf A}=A).
\]
By $\Delta_1(A)$ (and Markov's inequality again), there are at least 
$\frac{\eps}4s$ indices $j$, $1\leq j\leq s$, satisfying 
\begin{equation}\label{e10}
Pr(\B_j=\C'_j|\A=A)\geq 
\frac 12 +\frac{\eps}8.
\end{equation}
By $\Delta_2(A)$,
\[
\sum_j H(\C'_j|\A=A)\geq H(\C'|\A=A)=H(\C|\A=A)\geq n-\frac{4\xi}{\eps} s.
\]
Hence there are at least $\geq n-\frac{8\xi}{\eps} s$ indices $j$,
$1\leq j\leq n$ such that
\begin{equation}\label{e20}
H(\C'_j|\A=A)\geq 1/2.
\end{equation} 
Thus there are at least $\frac{\eps}4s - \frac{8\xi}{\eps} s$ indices $j$,
$1\leq j\leq s$ that satisfy both (\ref{e10}) and (\ref{e20}).
Setting $\xi=\frac{\eps^2}{64}$, this number  is
$\frac{\eps}8 s$. 
For such indices $I({\bf B}_j;{\bf C'}_j|{\bf A}=A)\geq \delta$, where
$\delta>0$ depends only on $\eps$, whence
\[
I({\bf B;C'|A}=A)\geq \frac{\eps\delta}8 s.
\]
By (\ref{e-b}) and (\ref{e-a}) we have
\begin{equation}\label{e3}
I({\bf A,B;C})\geq  \frac{\eps^2\delta}{32} s\geq 
\eta t^{1-\frac{1}{2k+1}}/k.
\end{equation}
where $\eta=\frac{\eps^2\delta}{32\cdot 360}$.
To get our lower bound on the communication complexity, we shall
compare the bounds (\ref{e0}) and (\ref{e3}).
Recall that $w$ consists of $2^{(k-2)r}$ messages, each having
length at most $(\log n)^C\tm l$, for some constant $C$.  Thus  
\[
I({\bf A,B;C})\leq |{\bf W}|\leq 
2^{(k-2)r}(\log n)^C\tm l=
2^{(k-2)\frac{\log t}{k-1}}(\log n)^C\tm l=
\]\[
t^{\frac{k-2}{k-1}}(\log n)^C\tm l\leq 
t^{1-\frac{1}{k-1}}(\log n)^C\tm l.
\]
Comparing this with the lower bound (\ref{e3}), we get
\[
l \geq  \fr{\eta t^{\fr{1}{k-1}-\fr 1{2k+1}}}{k\tm(\log n)^C}
\ge\fr{\eta n^{\frac 1{3k}\llp\fr1{k-1}-\fr 1{2k+1}\rrp}}{k\tm(\log n)^C}
\geq \fr{\eta n^{\frac 1{6k^2}}}{k\tm(\log n)^C}
\geq {n^{\gamma/k^2}},
\]
for a sufficiently small constant $\gamma$, as $k^2\in\aso{\fr{\log n}{\log\log n}}$.}

The theorem together with the protocol given in \prpref{pro_prot} for \prpref{pro_prot} leads to the following corollary:

\crl[cl_fin]{For $k(n)=\aso{\sqrt{\fr{\log n}{\log\log n}}}$, there exists a \f k-party relational communication problem that can be solved exactly by a quantum simultaneous message passing protocol of cost \asO{\log n} and requires superpolynomially more expensive protocols in the model of probabilistic non-interactive one-way communication with public randomness.
For $k=\asO1$ we get an exponential gap.}

\sect[sec_opr]{Open problems}
\itemi{ \item Extend the statement of \crlref{cl_fin} to bigger values of $k$.
\item Give a separation similar to the one shown here through a (partial) function.
How much can be saved by using quantum communication for total functions?
\item Give a separation for the multi-party interactive setting.
Even the case of three players and one-way interactive message passing (i.e., Alice speaks to Bob, after that Bob speaks to Charlie, who, in turn, responds) looks very interesting.}

\subsection*{Acknowledgments}
We would like to thank Michal Kouck\'{y} and Ji\v{r}\'{\i} Sgall for useful discussions about the problem considered in this paper.

Part of this work was done while D.~Gavinsky was visiting the
Institute of Mathematics, CAS in Prague.

\bib

\end{document}